\documentclass[a4paper,11pt,oneside,final]{article}

\usepackage[utf8]{inputenc}
\usepackage[T1]{fontenc}
\usepackage{amsmath}
\usepackage{amsfonts}
\usepackage{amssymb}
\usepackage{graphicx}
\usepackage{hyperref}
\usepackage[top=2.5cm, bottom=2.5cm, left=2.5cm, right=2.5cm]{geometry}

\title{Etching of semiconductors and metals by the Photonic Jet with shaped optical fiber tips}

\author{Robin Pierron\footnotemark[2] , Sylvain Lecler\footnotemark[2] \footnotemark[1] , Julien Zelgowski\footnotemark[2] , Pierre Pfeiffer\footnotemark[2] , Frédéric Mermet\footnotemark[3] , \\ Joël Fontaine\footnotemark[2]}

\date{Article published in \textit{Applied Surface Science} 418 (2017) \\ 
https://doi.org/10.1016/j.apsusc.2017.01.277}

\begin{document}

\maketitle

\footnotetext[2]{ICube Laboratory, CNRS, University of Strasbourg, France}
\footnotetext[3]{IREPA LASER – Carnot MICA Institute, Pôle API, 67412 Illkirch, France}
\footnotetext[1]{Corresponding author: sylvain.lecler@unistra.fr}



%
%
%
%
%
%
%

\begin{abstract}
The etching of semiconductors and metals by a photonic jet (PJ) generated with a shaped optical fiber tip is studied. Etched marks with a diameter of 1 micron have been realized on silicon, stainless steel and titanium with a 35 kHz pulsed laser, emitting 100 ns pulses at 1064 nm. The selection criteria of the fiber and its tip are discussed. We show that a 100/140 silica fiber is a good compromise which takes into account the injection, the working distance and the energy coupled in the higher--order modes. The energy balance is performed on the basis of the known ablation threshold of the material. Finally, the dependence between the etching depth and the number of pulses is studied. Saturation is observed probably due to a redeposition of the etched material, showing that a higher pulse energy is required for deeper etchings. 
\end{abstract}
%
%


\section{Introduction}

Direct laser etching at sub--wavelength scale remains a challenging issue. Numerical simulations have proven the feasibility of light concentration in the near field with a full width at half maximum (FWHM) smaller than the diffraction limit ($\lambda/2$) remaining smaller than $\lambda$ over a distance of few wavelengths depending on the conditions \cite{Lecler,Chen01}; the intensity can be as high as two hundred times the one of the incident wave \cite{Badreddine,Andri}. This phenomenon is called "Photonic Jet" (PJ). It can be obtained at the output of dielectric cylinders \cite{Chen01}, spherical particles \cite{Lecler}, square particles \cite{Pacheco} or planar waveguides \cite{Badreddine}. 

It has been shown that PJ can be used to downsize laser etching. Microspheres are deposited randomly directly on the sample \cite{Andri}, or periodically organized in a monolayer \cite{Grojo}. Experimental demonstrations have been performed at different wavelengths and pulse durations \cite{Andri, Munzer, Huang,Guo, Wu}. However, the particles are not easy to manipulate and they are polluted after the first etching due to their contact with the sample. One solution is to trap the sphere by an optical tweezer \cite{McLeod}; an another one that we propose, is to achieve the photonic jet using an optical fiber with a shaped tip, without spheres \cite{Julien}.

In our case, the shaped tip has a given length and a shape described thanks to a rational B\'ezier curve having a weight $w$ \cite{Julien}. Several tip shapes have been investigated: from triangular ($w=0$) to quasi elliptical ($w\simeq0.7$).

In this paper, a systematic study of the metal etching using a near--infrared (NIR) nanosecond pulsed photonic jet at the end of shaped fiber tips is presented. Several optical fibers with shaped tips have been experimented. The influence of the core size and the tip shape on the working distance, on the etching size and on the energy balance is discussed. The selected fiber tip has been tested on silicon, stainless steel and titanium. The required injected power is discussed in relation with the known ablation threshold. Finally, the influence of the number of pulses on the ablation depth is studied at the minimum etching size. 

\section{Experimental details}

The experimental setup consists in a NIR pulsed laser (VGEN ISP 1--40--30) commonly used for industrial laser processing, a focusing lens (achromatic doublet Thorlabs AC127--019--C), a XYZ stage with FC or SMA connectors, an optical fiber on a motorized Z--axis stage and a sample target placed perpendicularly to the shaped fiber tip on a motorized X--Y stage. The experiments are carried out in ambient atmosphere conditions. The laser works at a central wavelength of 1064 nm, a pulse duration of 100 ns and a repetition rate of 35 kHz. The setup is controlled by LabView. 

Three different step--index silica fibers were used: (1) a single mode 20/125 Large Mode Area (LMA) with a numerical aperture (NA) of 0.08; (2) a multimode 50/125, NA = 0.22; (3) a multimode 100/140, NA = 0.22. The distance between the fiber tip and the sample is controlled by a camera with a telecentric objective (5x magnification). 
The optical fiber tip shape is described using a Bezier curve with a weight $w$ chosen to give a PJ with a FWHM of 1 $\mu$m, located a few tens micrometers from the tip.

Three types of materials were tested: monocrystalline silicon (wafer) with a passive layer and an initial roughness ($R_a$) of 2 $\pm$ 1 nm, stainless steel ($R_a$ = 187 $\pm$ 14 nm) and titanium ($R_a$ = 423 $\pm$ 20 nm). No additional mechanical polishing was performed leaving relatively significant striations on the stainless steel and titanium samples. For the characterization after laser ablations, two different optical methods have been used: an optical microscope with a lateral resolution of 610 nm (50x focusing objective Zeiss, NA = 0.55) and a Zygo NewView 7200 profilometer with an axial resolution of 10 nm and a lateral resolution of 550 nm (50x Mirau objective, NA = 0.55). The power at laser and fiber outputs has been measured with a calorimeter Ophir Vega with a 3A--P cell.

\section{Results and discussion}

	\subsection{Influence of the fiber core diameter}

\begin{figure}[!ht]
	\centering
	\includegraphics[scale=0.3]{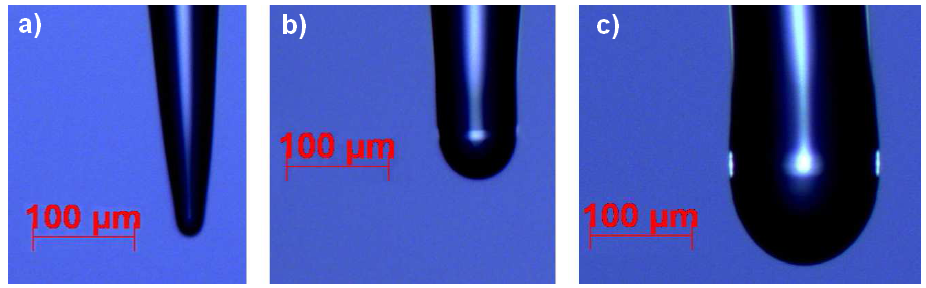}
	\caption{Shaped optical fiber tips designed to achieve PJ with a FWHM of 1 $\mu m$. (a) LMA 20/125 fiber (b) 50/125 fiber and (c) 100/140 fiber.}
	\label{fig:fibertips}
\end{figure}

Theoretically, it is possible to achieve PJ having a FWHM smaller than a half wavelength \cite{Lecler,Chen01}. However previous studies have shown that, in this case, the PJ intensity maximum was just on the tip end \cite{Julien}, which was not a convenient situation. Therefore, the first step has been to study the influence of the core diameter size of the fiber on the PJ position in order to select a fiber achieving a PJ with a minimum FWHM around 1 $\mu m$, at a distance from the tip large enough to avoid any disturbance. 
The fiber tips have been achieved by the LovaLite company thanks to thermoforming. They are shown in Figure \ref{fig:fibertips}. The tests have been performed on the wafer of silicon at the minimum power to ablate. The number of pulses was 35 for each PJ etching.

The fiber tip parameters and the results are summarized in Table \ref{tab1:core_diameter}. The PJ$_{distance}$ is the distance between the end of the tip and the maximum intensity in the PJ. It is the working distance in our etching process. As expected, the diameters of PJ etching marks (PJ$_{mark}$) are around 1.0 $\pm$ 0.6 $\mu m$ whatever the core diameter, which corroborates our numerical simulations. We can deduce that the size of PJ marks does not depend on fiber core diameter but on the shaped fiber tip (L$_{tip}$, $w$).

\begin{table}[!h]
\centering
\begin{tabular}{|c|c|c|c|c|c|c|}
\hline Fiber & L$_{tip}$ [$\mu m$] & $w$ & Injection [\%] &  PJ $_{distance}$ [$\mu m$] & PJ$_{mark}$ [$\mu m$] & E$_{pulse}$ [$\mu$J] \\ 
\hline 20/125 & 5 $\pm$ 1& 0.4 & 9 $\pm$ 3& 15 $\pm$ 2 & 1.0 $\pm$ 0.6 & 16 $\pm$ 5\\
\hline 50/125 & 17 $\pm$ 1 & 0.5 & 38 $\pm$ 4 & 51 $\pm$ 2 & 0.8 $\pm$ 0.6 & 7 $\pm$ 1\\
\hline 100/140 & 74 $\pm$ 1 & 1.0 & 46 $\pm$ 5 & 97 $\pm$ 2 & 1.2 $\pm$ 0.6 & 34 $\pm$ 4\\
\hline
\end{tabular}
\caption{Dimensions of the optical fiber; tip parameters numerically designed to achieve PJ with FWHM of 1 $\mu m$; corresponding minimum pulse energy injected to begin to etch silicon; associated PJ distances and mark diameters. 35 pulses were used for each PJ etching.}
\label{tab1:core_diameter}
\end{table}

Besides, the results show that to get a PJ with a FWHM of 1 $\mu m$, the corresponding PJ distance must be of the same order than the fiber core diameter (relative deviation less 30\%). In order to avoid the possible deposition of ablated material on the fiber tip, the fiber of 100 $\mu m$ core was selected; its working distance is 97 $\pm$ 2 $\mu m$.

	\subsection{Influence of the optical fiber on the minimum required power}

The 1 $\mu m$ etching can be achieved independently on the fiber core diameter, however the minimum required injected power is not the same. For example, the results (cf. Table \ref{tab1:core_diameter}) show that the pulse energy ($E_{pulse}$) injected in the 100 microns core fiber and necessary to alter the material is five times higher than the one injected in the 50 microns core diameter. 
Using a shaped optical fiber tip, the photonic jet is due to the fundamental mode. The higher groups of modes are off--axis focused on larger surfaces, giving lower power densities, insufficient to etch. The number of mode groups increases with the size of the fiber core \cite{Snyder}. The injected energy is then spread in these groups and the intensity in the fundamental mode becomes lower. 
Therefore, a singlemode fiber seems to have an interest. The 20/125 LMA fiber is singlemode. However, because of the smaller core diameter, the injection rate is four times
smaller than for the two other fibers (cf. Table \ref{tab1:core_diameter}). Thus, to select the fiber core diameter, a compromise must be found between the working distance, the injection condition and the required etching power.

\begin{figure}[!h]
	\centering
	\includegraphics[scale=0.4]{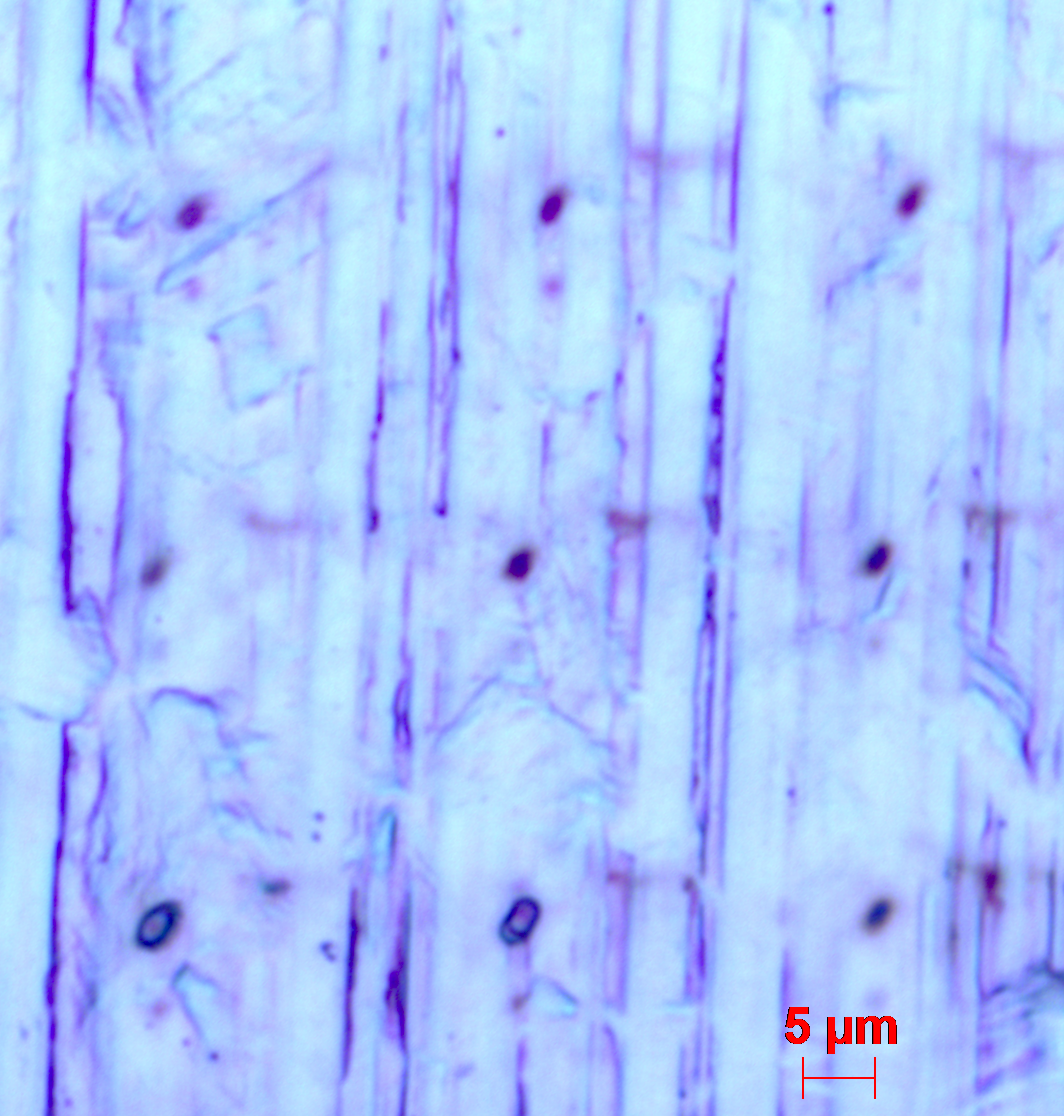}
	\caption{Etched marks with a diameter of around 0.8 $\pm$ 0.6 $\mu$m on the stainless steel. 100/140 fiber with tip numerically designed to achieve the PJ with FWHM of 1 $\mu m$. Ablations were done at 3 $\pm$ 1 $\mu$J pulse energy corresponding to the minimum to etch. 35 pulses were used for each PJ etching.}
	\label{fig:matrix}
\end{figure}

In the following part, etchings on stainless steel and titanium with a 100/140 fiber have been studied. With this fiber, the injection efficiency is higher than the one of the LMA fiber and the working distance is larger than the one of the 50/125 fiber. On the two metals, etching marks around 1.1 $\mu m$ have been successful achieved without polluting the fiber tip. The results (cf. Table \ref{tab2:PJ_metal}) show that a similar minimum energy per pulse (E$_{pulse}$) is required to etch the two materials. The PJ mark diameters (PJ$_{mark}$) have the same size order, almost 1.2 $\mu m$ on titanium and stainless steel (cf. Figure \ref{fig:matrix}). 

For comparison, the theoretical pulse energy required to reach the ablation threshold with a singlemode fiber (E$_{pulse}$ SM) has been computed. In this case all the energy is assumed to be concentrated on the etched surface (values Table \ref{tab2:PJ_metal}). Under our operating conditions (1064 nm, 100 ns, 35 kHz), a threshold fluence of $\sim$4.5 J/cm$^2$ for titanium \cite{Vladoiu} and  $\sim$4.0 J/cm$^2$ for stainless steel have been considered. The required energy with the 100 $\mu m$ fiber is almost 100 times larger than the one considered in a singlemode. This factor is in accordance with the estimation of the mode groups number M=46 ($M=2\pi/\lambda \cdot NA \cdot a /\sqrt{2}$ where $a$ is the fiber radius), taking into account that knowing the exact energy distribution on the guided modes is difficult.

\begin{table}[!ht]
\centering
\begin{tabular}{|c|c|c|c|}
\hline Metal &  PJ$_{mark}$ [$\mu m$] & E$_{pulse}$ [$\mu$J] & E$_{pulse}$ SM [$\mu$J] \\ 
\hline Stainless steel & 0.8 $\pm$ 0.6 & 3 $\pm$ 1 & 0.020 $\pm$ 0.001 \\
\hline Titanium & 1.4 $\pm$ 0.6 & 4 $\pm$ 1 & 0.069 $\pm$ 0.004 \\
\hline
\end{tabular}
\caption{For stainless steel and titanium: PJ marks diameters (PJ$_{mark}$), minimum pulse energy injected to begin to etch (E$_{pulse}$) and theoretical energy by pulse injected to etch if all the energy would be concentrated on the etched surface. 100/140 fiber with tip numerically designed to achieve PJ with FWHM of 1 $\mu m$. 35 pulses were used for each PJ etching.}
\label{tab2:PJ_metal}
\end{table}

	\subsection{Influence of the number of pulses on the ablation depth}

Finally, the relation between the number of pulses and the ablation depth has been studied for the parameters giving the minimum etching size. The experiments have been performed with the 100/140 fiber on the silicon wafer and measured with the Zygo profilometer. The pulse energy was constant at 34 $\pm$ 4 $\mu$J corresponding to the minimum injected energy required to ablate. In order to validate the repeatability, three series of etchings were produced for each value of pulse number. 

\begin{figure}[!h]
	\centering
	\includegraphics[scale=1]{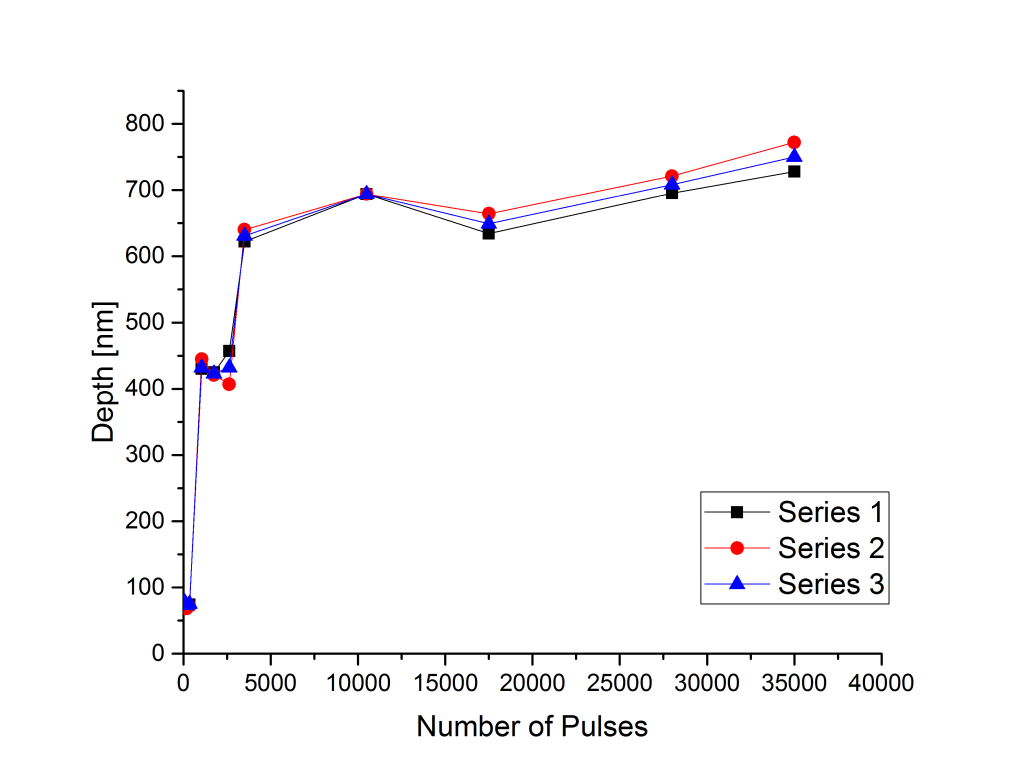}
	\caption{The ablation depth [nm] as a function of the number of laser pulses. 100/140 fiber with tip numerically designed to achieve the PJ with FWHM of 1 $\mu m$. Ablations were done at 34 $\pm$ 4 $\mu$J pulse energy corresponding to the minimum to etch.}
	\label{fig:Depth_vs_pulses}
\end{figure}
	
Figure \ref{fig:Depth_vs_pulses} shows the ablation depth versus the number of laser pulses for the silicon wafer. The ablation depth increases significantly during the first 3500 laser pulses. Then, the depth reaches a maximum value around 700 nm. The fiber tip is not polluted and can still achieve new etchings. The computed photonic jet field depth reaches a few tenths of microns, which can not be the explanation. We suppose that the etched matter is probably redeposited because of the low energy of each pulse.

\begin{figure}[!h]
				\centering
				\includegraphics[scale=1]{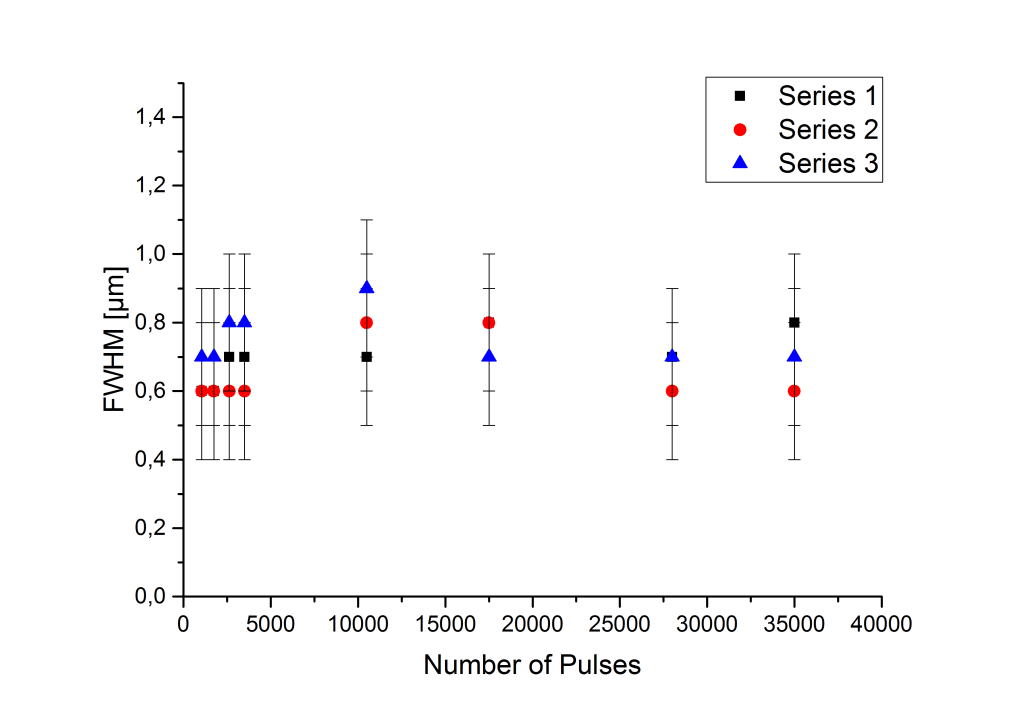}
				\caption{The etched mark FWHM [$\mu$m] as a function of the number of laser pulses. 100/140 fiber with tip numerically designed to achieve the PJ with FWHM of 1 $\mu m$. Ablations were done at 34 $\pm$ 4 $\mu$J pulse energy corresponding to the minimum to etch.}
				\label{fig:fwhm_vs_pulses}
\end{figure}

\begin{figure}[!h]
	\centering
	\includegraphics[scale=1]{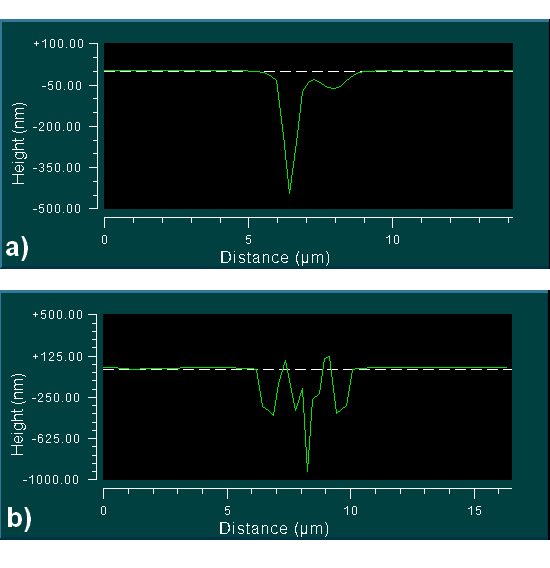}
	\caption{Examples of ablation profiles on silicon for (a) 1050 pulses and (b) 17500 pulses. 100/140 fiber. Ablations were done at 34 $\pm$ 4 $\mu$J.}
				\label{fig:profil}
\end{figure}

Besides, the etched mark FWHM as a function of the number of pulses have been measured (cf. Figure \ref{fig:fwhm_vs_pulses}). The  FWHM is constant and measures almost 0.8 $\pm$ 0.6 $\mu m$ independently of the number of pulses, which was not expected. Generally the etched mark diameters have a logarithmic dependence with the numbers of pulses \cite{Mannion}. The fact that the etching depth does not increase anymore can be a reason. Moreover, as illustred in Figure \ref{fig:profil}, we see that the complexity of etched marks profiles increases with the number of pulses probably due to the higher--order modes contribution. Higher pulse energies may be required for deeper etchings but likely with an increase of the etched mark diameters. 

\section{Conclusions}

An experimental study of the influence of the fiber core diameter and the tip shape, on the working distance and the minimum required energy to etch has been performed on a silicon wafer with a nanosecond infrared laser (35 laser pulses, 1064 nm, 100 ns). On this basis, we showed that a compromise between the working distance, the injection rate and the lost energy in the higher guided modes led to choose the 100/140 shaped tip fiber. The photonic jet maximum intensity is located at almost 100 $\mu m$ of fiber tip. With this fiber, etching diameters of around 1.1 $\mu m$ have been demonstrated on silicon, stainless steel and titanium. On silicon with a pulse energy as small as 34 $\pm$ 4 $\mu$J, after a first regime where the depth increases with the pulse number, a maximum is reached at around 700 nm ($<1 \mu m$). It is probably due to a material redeposition owing to the low pulse energy, suggesting that a higher pulse energy is required for deeper etchings. 

\section*{Acknowledgements}

We would like to thank Audrey Leong--Ho\"i (ICube Laboratory, France) for the measures of roughness of metal samples with the Leiz--Linnik microscope developed at ICube (50x objective, NA = 0.85). 
We acknowledge also the financial support from Conectus Alsace, a member of SATT network.


\bibliographystyle{ieeetr} 
\bibliography{mybibfile}

\begin{thebibliography}{10}

\bibitem{Lecler}
S.~Lecler, Y.~Takakura, and P.~Meyrueis, ``Properties of a three-dimensional
  photonic jet,'' {\em Opt. Lett.}, vol.~30, pp.~2641--2643, Oct 2005.

\bibitem{Chen01}
Z.~Chen, A.~Taflove, and V.~Backman, ``Photonic nanojet enhancement of
  backscattering of light by nanoparticles: a potential novel visible-light
  ultramicroscopy technique,'' {\em Opt. Express}, vol.~12, pp.~1214--1220, Avr
  2004.

\bibitem{Badreddine}
B.~Ounnas, B.~Sauviac, Y.~Takakura, S.~Lecler, B.~Bayard, and S.~Robert,
  ``Single and dual photonic jets and corresponding backscattering enhancement
  with tipped waveguides: Direct observation at microwave frequencies,'' {\em
  IEEE Transactions on Antennas and Propagation}, vol.~63, pp.~5612--5618, Dec
  2015.

\bibitem{Andri}
A.~Abdurrochman, S.~Lecler, F.~Mermet, B.~Y. Tumbelaka, B.~Serio, and
  J.~Fontaine, ``Photonic jet breakthrough for direct laser microetching using
  nanosecond near-infrared laser,'' {\em Appl. Opt.}, vol.~53, pp.~7202--7207,
  Nov 2014.

\bibitem{Pacheco}
V.~Pacheco-Pena, M.~Beruete, I.~V. Minin, and O.~V. Minin, ``Multifrequency
  focusing and wide angular scanning of terajets,'' {\em Opt. Lett.}, vol.~40,
  pp.~245--248, Jan 2015.

\bibitem{Grojo}
D.~Grojo, L.~Charmasson, A.~Pereira, M.~Sentis, and P.~Delaporte, ``Monitoring
  photonic nanojets from microsphere arrays by femtosecond laser ablation of
  thin films,'' {\em Journal of Nanoscience and Nanotechnology}, vol.~11,
  no.~10, pp.~9129--9135, 2011.

\bibitem{Munzer}
H.-J. M\"{u}nzer, M.~Mosbacher, M.~Bertsch, J.~Zimmermann, P.~Leiderer, and
  J.~Boneberg, ``Local field enhancement effects for nanostructuring of
  surfaces,'' {\em Journal of Microscopy}, vol.~202, no.~1, pp.~129--135, 2001.

\bibitem{Huang}
S.~M. Huang, M.~H. Hong, B.~S. Luk’yanchuk, Y.~W. Zheng, W.~D. Song, Y.~F.
  Lu, and T.~C. Chong, ``Pulsed laser-assisted surface structuring with optical
  near-field enhanced effects,'' {\em Journal of Applied Physics}, vol.~92,
  no.~5, pp.~2495--2500, 2002.

\bibitem{Guo}
W.~Guo, Z.~B. Wang, L.~Li, D.~J. Whitehead, B.~S. Luk’yanchuk, and Z.~Liu,
  ``Near-field laser parallel nanofabrication of arbitrary-shaped patterns,''
  {\em Applied Physics Letters}, vol.~90, no.~24, p.~243101, 2007.

\bibitem{Wu}
W.~Wu, A.~Katsnelson, O.~G. Memis, and H.~Mohseni, ``A deep sub-wavelength
  process for the formation of highly uniform arrays of nanoholes and
  nanopillars,'' {\em Nanotechnology}, vol.~18, no.~48, p.~485302, 2007.

\bibitem{McLeod}
E.~Mcleod and C.~B. Arnold, ``Subwavelength direct-write nanopatterning using
  optically trapped microspheres,'' {\em Nature Nanotechnology}, vol.~3,
  pp.~413--417, July 2008.

\bibitem{Julien}
J.~Zelgowski, A.~Abdurrochman, F.~Mermet, P.~Pfeiffer, J.~Fontaine, and
  S.~Lecler, ``Photonic jet subwavelength etching using a shaped optical fiber
  tip,'' {\em Opt. Lett.}, vol.~41, pp.~2073--2076, May 2016.

\bibitem{Snyder}
A.~W. Snyder and J.~D. Love, {\em Optical waveguide theory}.
\newblock Springer, 1$^{st}$~ed., 1983.
\newblock p. 704--706.

\bibitem{Vladoiu}
I.~Vladoiu, M.~Stafe, C.~Negutu, and I.~M. Popescu, ``Nanopulsed ablation rate
  of metals dependence on the laser fluence and wavelength in atmospheric
  air,'' {\em UPB Scientific Bulletin, Series A: Applied Mathematics and
  Physics}, vol.~70, no.~4, pp.~119--126, 2008.

\bibitem{Mannion}
P.~Mannion, J.~Magee, E.~Coyne, G.~O’Connor, and T.~Glynn, ``The effect of
  damage accumulation behaviour on ablation thresholds and damage morphology in
  ultrafast laser micro-machining of common metals in air,'' {\em Applied
  Surface Science}, vol.~233, no.~1–4, pp.~275 -- 287, 2004.

\end{thebibliography}

\end{document}